\documentstyle[epsf]{mn}
\title{Gravitational instability and star formation in disk galaxies}
\author[U. S. Pandey \& C. van de Bruck]{U. S. Pandey$^{1,2}$ \& C. 
van de Bruck$^{1}$\\
$^1$Institut f\"ur Astrophysik der Universit\"at Bonn, 
Auf dem H\"ugel 71, D-53121 Bonn, Germany\\
$^2$Department of Physics, DDU Gorakhpur University, Gorakhpur -- 273009, 
India}
\begin{document}
\maketitle
\begin{abstract}
We present a general star formation law where star formation rate depends 
upon efficiency $\alpha$, timescale $\tau$ of 
star formation, gas component $\sigma_{g}$ of surface mass density and a 
real exponent $n$. A given exponent $n$ determines 
$\tau$ which however yields the corresponding star formation rate. Current 
nominal Schmidt exponent $n_{s}$ for our 
model is $2<n_{s}<3$. Based on a gravitational instability parameter 
$Q_{A}$ and another dimensionless 
parameter $f_{P}\equiv (P/G\sigma_{c}^{2})^{1/2}$, where $P$ = pressure, 
$\sigma_{c}$ = column density of molecular clouds, we suggest a 
general equation for star formation rate which depends upon relative 
competence of the two parameters for various physical circumstances. 
We find that $Q_{A}$ emerges to be a better parameter for star formation 
scenario than Toomre Q-parameter. Star formation rate in the 
solar neighbourhood is found to be in good agreement with values 
inferred from previous studies. Under closed box approximation model, 
we obtain a relation between metallicity of gas and the efficiency of 
star formation. Our model calculations of metallicity in the solar 
neighbourhood agree with earlier estimates. We conclude that metallicity 
dispersion for stars of same age may result due to a change in 
efficiency through which different sample stars were processed. For no 
significant change of metallicity with age, we suggest that all sample 
stars were born with almost similar efficiency.
\end{abstract}

\section{Introduction}
It was realized by Kenicutt (1989) that there occurs non-linear increase in
the star formation rate near the threshold surface density corresponding to
$Q$-parameter. This shows in fact why rate of star formation $R$ is large in 
the spiral arms of several galaxies. For example, in M 51 and NGC 6946 
(Rydbeck, Hjalmarson \& Rydbeck 1985; Lord 1987; Tacconi-Garman 1988), 
gas densities in
the spiral arms are larger by a factor of two indicating deviations in the
usual power law exponent ($n\gg 2$) of Schmidt (1959, 1963). In fact, 
star formation in many spiral galaxies under extreme conditions of low 
gas density and low disk self--gravity present a challange to all current 
theories for disk star formation (Ferguson et al. 1996). New star formation 
laws have therefore been proposed (see e.g. Talbot \& Arnett 1975; Dopita
1985; Wyse 1986; Silk 1987; K\"oppen \& Fr\"ohlich 1997). 

However, a global star formation law has been
put to doubt (Hunter \& Gallagher 1986). For a general star formation scenario
one may refer to Zinnecker \& Tscharnuter (1984); Zinnecker (1989 and 
references 
therein). Thus, many interesting variants on the simple star formation laws
include e.g. self-propagating star formation (stochastic) in 
Gerola \& Seiden (1978); 
Seiden \& Gerola (1982); Dopita (1985); Coziol (1996); 
self-propagating star formation 
in Arimoto (1989) and Hensler \& Burkert (1990a, 1990b); star formation 
bursts (stochastic) in Matteucci \& Tossi (1985). Kr\"ugel \& Tutukov (1993) 
and Tutukov \& Kr\"ugel (1995) have used one--zone dynamical code without 
radial dependence 
of the variables to study the conditions for bursts of star formation. In the later 
paper, using one--zone code, they studied other types of burst of star formation in a 
galactic nucleus different from periodic bursts. Further, surface gas density threshold 
for star formation has been discussed in Kenicutt (1989).

Wyse \& Silk (1989) have discussed an extended Schmidt model with $R$ 
dependence on surface gas density $\sigma_{g}$ and local angular frequency 
$\Omega(r)$ for both atomic and molecular gases respectively with $n = 1$ and 
$n = 2$. Wang \& Silk (1994) have presented recently a self-consistent 
model (considering total gas surface density) for global star formation 
based on the 
gravitational instability parameter $Q<1$ by Toomre (1964). In the solar 
neighbourhood, 
the model agrees with (i) the observed star formation rate, (ii) the 
metallicity distribution among G-dwarfs, and (iii) the age metallicity 
relation 
for F-dwarfs. The model may be compared for star formation rate in galactic 
disks with Schmidt law with an exponent of about 2. Star formation rate 
depends 
also on the epicyclic frequency. A natural cut-off for $Q = 1$ in the star 
formation rate results. However, their analysis is heavily based on $Q<1$ 
criterion 
which has been put to question in relation to non-radial instabilities 
in the galactic disks that may play more fundamental role when magnetic field 
supported by azimuthal gas motions (thus resulting thermal instability not 
at all related to Q) is taken into consideration (Elmegreen 1993). We note 
(see 
e.g. Figures 4 and 6 of Wang \& Silk 1994) that the star formation does 
proceed 
in the regions where $Q\geq 1$. A natural question to ask is: how does 
star formation 
occur when $Q\geq 1$ and consequently the system has attained the state of 
gravitational 
equilibrium? We attempt here to precisely answer this question and provide 
a scenario to 
circumvent this natural cut-off in the star formation process (see Sect. 3 
in the text 
for details). The fact that star formation occurs via gravitational 
instability 
was also suggested by Fall \& Efstathiou (1980). The Q-regulation near its 
threshold value has been discussed by Dopita (1985) and Silk (1992). Silk 
(1995) has argued that local self--regulation of star formation may help 
expain the initial mass function of stars and that global self--regulation 
can account for the rate of star formation. Effect of environment on the 
gas content and rotation curves of disk may play a crucial role in 
determining star formation rates and histories. 

Elmegreen (1995) has discussed critical column densities for gravitational 
instabilities and for cooling to diffuse cloud temperatures. It has been 
shown that the fundamental scale for star formation in the outer regions 
of galaxies (in spiral arms) and in the resonance rings are related to 
the local unstable length. Since critical gas density for gravitational 
instability scales as local density, inner regions of galaxies have higher 
star formation rate beyond threshold density.

The consideration of magnetic field changes velocity dispersion
by a factor of two pushing $Q>1$, (i.e. stable region). Incorporating this with
the fact that there occurs shear instability of magnetised gas in the azimuthal
direction, one is stimulated to think that $Q<1$ may not be the only criterion
for cloud formation which leads to star formation. An alternative suggestion 
for cloud formation because of energy dissipation accompanied by shear instability 
thus leading to star formation (even if $Q>1$) has been given (Elmegreen 1991a,
1993, and below for details). Macroscopic thermal instabilities and various cloud formation 
mechanisms are reviewed in Elmegreen (1991b). We assume that instability parameter suggested by
\mbox{Elmegreen (1993)}, i.e. $Q_{A}<1$ (instead of $Q<1$) is the criterion which determines 
occurrence of significant cloud formation instabilities. A natural consequence 
of our analysis is that star formation proceeds in the regions where one
has $Q\geq 1$. It may be noted that in these regions (system being gravitationally 
stable) an altogether different cloud formation mechanism (leading to star 
formation) as suggested by Elmegreen (1993) is asked for.  We shall present
subsequently the evidence in support of our assumption. The outline of the
paper is as follows: we give a general law for star formation rate in Section 2. In 
Section 3, we suggest a general equation for star formation rate which depends 
upon two fundamental parameters $Q_{A}$ and $f_{P}$ (defined in the text). We also 
give  a comparison of star formation rate in the solar neighbourhood and timescale 
of gas depletion. Variations in the star formation rate and metallicity 
distribution in the solar neighbourhood are discussed in Section 4. The Section 
5 gives discussions and a resume of our results.

\section{STAR FORMATION RATE}
 
We write the star formation rate in the form 
\begin{equation}
R^{n} = \alpha (\sigma_{g}/\tau)^{n}    
\end{equation}
where $\alpha$ = efficiency of star formation which also depends upon $n$, 
$\tau$ = timescale of star formation, $\sigma_{g}$ = surface density of gas
composed of atomic and molecular components, \mbox{$n$ = an exponent}. Clearly, 
$\tau^{-1}$ is related to growth rate of instability (Goldreich \& Lynden-
Bell 1965). A review of recent observations of the history of star 
formation and its relevance to galaxy formation and evolution has been 
discussed by Kennicutt (1996). For the evolution of the global star 
formation history measured from the Hubble Deep Field one may refer 
to Connolly et al. (1997). 
Gravi\-tational instability of galactic disks has also been studied
by \mbox{Elmegreen (1979)}, Cowie (1981), Ikeuchi, Habe \& Tanaka (1984) and 
Bizyaev (1997). However, gravitational instabilities in the presence 
of turbulence are discussed in Bonazzola et al. (1987) and 
Leorat, Passot \& Pouquet (1990). It is found that supersonic turbulence may 
be strong enough (in some 
cases) to hold the Jeans criterion for gravitational instability. As a result, 
it may stop gravitational 
collapse. In this scenario, star formation takes place in molecular cloud 
complexes at places where the turbulence evolves to 
subsonic phase. In the present analysis we do not aim to discuss the instability 
criteria and their relevance to star formation 
(which are certainly interesting topics of research at present), instead we aim 
to obtain a general star 
formation law with small number of adjustable parameters.
We assume neither infall nor radial flow in the disk. We consider 
gravitational instability due to axisymmetric perturbations (for non-axisymmetric case, 
one may refer to Goldreich \& Lynden-Bell 1965) with magnetic field in the azimuthal 
direction which gives rise to shear instability in a magnetised gas. 
Groth rate of instability is now expressed as
\begin{equation}
\omega^{2} = k^{2}v_{eff}^{2} - 2\pi G\sigma_{g}k + \kappa^{2}
\end{equation} 
where $k$ is the wave number and $\kappa$ is the epicyclic 
frequency, $v_{eff}$ is the effective 
velocity dispersion for ambient Alfven speed such that
\begin{equation}
v_{eff} = (v^{2}\gamma_{eff} + v_{Alf}^{2})^{1/2},
\end{equation}
$v$ being the velocity dispersion without magnetic field and
\begin{equation}
\gamma_{eff}=\frac{\gamma\omega -\omega_{c}(1+s-2r)}{\omega+\omega_{c}(3-s)}.
\end{equation}
Here $\gamma$ = ratio of two specific heats, $\omega_{c}$ = cooling rate 
(see e.g. Elmegreen 1993 for details).
The parameter $Q$ is written as $Q \equiv \kappa v_{eff}/\pi G \sigma_{g}$. 
Gravitational instability requires that both $Q<1$ and $k$ are smaller than a critical value
\begin{equation}
k_{cr}=\frac{\pi G \sigma_{g}}{v_{eff}^{2}}\left(1+\left(1-Q^{2}\right)^{1/2}\right).
\end{equation}
Due to thermal instability, if $\gamma_{eff}$ reaches large negative values
(such that $v_{eff}^{2}<0$), it implies no critical (or minimum) wavelength 
for gravitational perturbation in the radial direction. This makes $Q^{2}<0$. 
However, we do have a maximum wavelength of the perturbation. Thus, equation
(2) shows the absence of Q-threshold for azimuthal instability which means that 
all Q-values provide unstable growth. Q-threshold may appear only if 
$\gamma_{eff}(\omega)$ becomes a constant. Therefore, for the present treatment, 
we demand $Q_{A}\equiv 2\sqrt2 A v_{eff}/\pi G \sigma_{g}<1$ for growth of 
gravitational instability but we are well aware that thermal and shear instabilities  
(along azimuthal direction) are capable of determining cloud formation 
leading to star formation even if $Q>1$. This  amounts to replacing 
$\kappa$ by $2\sqrt2 A$ in the original $Q$ \mbox{(a scale transformation for Keplerian 
disk).}

Maximum of $\omega^{2}$ occurs at 
\begin{equation}
k_{max} = 2\sqrt2 A/v_{eff} Q_{A}
\end{equation}
($A$ = Oort shear constant), which provides maximum growth rate as (Wang \& Silk 1994)
\begin{equation}
|\omega_{max}| = \frac{2\sqrt{2} A(1-Q_{A}^{2})^{1/2}}{Q_{A}}.
\end{equation}
Since, $\tau \simeq |\omega_{max}|^{-1}$, one gets from equations (1) and (7)
\begin{equation}
R^{n}=\frac{\alpha(2\sqrt{2} A)^{n} \sigma_{g}^{n} (1 - Q_{A}^{2})^{n/2}}{Q_{A}^{n}}.
\end{equation}
Following Wang \& Silk (1994), we define a function $f_{c} = \sigma_{g}/ \sigma_{c}$, 
$\sigma_{c}$ = column density of individual molecular clouds. But, however,
the relation between individual cloud formation and star formation is
complicated. Even cloud formation process is not well known. The assumption
that star formation rate results due to gravitational instability naturally
demands for its relation with cloud formation process. Elmegreen (1990)
has shown that gravitational instabilities generally form giant molecular
clouds faster than due to random collisions. Cloud formation followed by
star formation in the interstellar medium is certainly not the purpose
of our investigation. Under the assumption that only gravitational 
instability is predominant, small cloud collisions may lead to large
molecular clouds wherein star formation ensues. It is then natural to think
that within an order of magnitude cloud formation timescale or equivalently
cloud collision timescale and growth rate of local instability timescale
are similar. With this scenario, Wang \& Silk (1994) derive the expression for 
collision time between two clouds. We, thus, make use of their result 
and write collision time 
between two clouds as    
\begin{equation}
t_{coll}^{-1} = \frac{\sigma_{g} (2\sqrt{2} A)}{\sigma_{c} Q_{A}}.
\end{equation}
In view of the above, $t_{coll}^{-1}\sim \omega_{max}$, and one gets
\begin{equation}
Q_{A} \sim (1-f_{c}^{2})^{1/2}.
\end{equation}
It is to be noted that this may not reflect the general property of the
interstallar medium, e.g. other types of instabilities namely thermal
instability and Parker instability might also contribute   
and affect the timescale of star formation (subsequently other physical
quantities).       
Making use of eq. (10) into eq. (8), star formation rate is now expressed as 
\begin{equation}
R^{n} = \frac{\alpha (2\sqrt{2} A )^{n} \sigma_{g}^{n} f_{c}^{n}}{(1-f_{c}^{2})^{n/2}}. 
\end{equation}
Eventually, in this form eq. (11) now assumes the conversion from
column density to density using the galactic scale height.
Let us write equation (11) in the form 
\begin{displaymath}
n\frac{\partial \ln R}{\partial \ln \sigma_g} = n + n\frac{\partial \ln A}{\partial \ln \sigma_g} + 
n\frac{\partial}{\partial \ln \sigma_g}\left[ \ln\left( \frac{f_c}{(1-f_c^2)^{1/2}}\right) \right]
\end{displaymath}
or
\begin{equation}
n_{s} \equiv \frac{\partial \ln R}{\partial \ln \sigma_{g}} = 2 + \frac{\partial \ln A}{\partial \ln \sigma_{g}} + \frac{f_{c}^{2}}{1-f_{c}^{2}},
\end{equation}
where $n_{s}$ stands for the nominal Schmidt exponent. The second term in 
equation (12) appears because for spiral waves epicyclic frequency is expressed through 
\begin{equation}
\kappa = \kappa_{0} (\sigma_{g}/\sigma_{0})^{1/2}
\end{equation}
and shear constant A is 
\begin{equation}
A = A_{0} (2 - \sigma_{g}/\sigma_{0}).     
\end{equation}
Non-axisymmetric gravitational perturbation of a magnetic gaseous disk has been 
discussed by Elmegreen (1987) who has obtained eq. (13) and eq. (14). Here, 
$A_{0}$ and $\sigma_{0}$ represent equilibrium values of the shear rate
and the surface mass density (see also  Waller \& Hodge 1991). 
It is easy to see that for 
vanishing shear constant, equation (12) reduces to equation (19) of Wang \& Silk
(1994, hereafter WS). It may be regarded as generalised version of WS equation 
in the sense that there is an additional term on the right side which is certainly 
non-zero. We calculate the second term on the right of equation (12), i.e., 
$\partial \log A/ \partial \log \sigma_{g} \sim 0.54$ using least squares method. 
Data reported in 
Table 1 have been taken from Einasto (1979) and  WS. Since mostly $f_{c}$ is 
very small compared to unity (see e.g. Table 5) for present Galactic disk, we 
conclude that the nominal Schmidt exponent $n_{s}$  for our model corresponds to 
$2<n_{s}<3$ for the Galaxy. For usual Schmidt law $n_{s}$ lies between 1 and 2. 
The other normal spiral galaxies of the Galaxy type are supposd to follow the same signature.

\begin{table}
\caption{Variation of shear constant $A$ with surface density}
\begin{center}
\begin{tabular}{ccccc}
\hline 
  Distance & A & log A & $\sigma _{g}$ & log$\sigma _{g}$ \\
  (kpc) & (km $s^{-1} kpc^{-1}$ ) &   & ($ M_{\odot} pc^{-2}$) &   \\
  \hline
1      &  105   &               2.0212   &100   &               2.0000\\
2      &   30   &               1.4771   &  3   &               0.4771\\
3      &   20.9 &               1.3202   &  5   &               0.6990\\
4      &   19.7 &               1.2945   & 10   &               1.0000\\
5      &   19.1 &               1.2820   & 10.5 &               1.0212\\
6      &   18.2 &               1.2601   & 10.2 &               1.0086\\
7      &   17.2 &               1.2355   & 10   &               1.0000\\
10     &   13.8 &               1.1399   &  7   &               0.8451\\
12     &   11.5 &               1.0607   &  5   &               0.6990\\
14     &    9.6 &               0.9823   &  4   &               0.6021\\
16     &    7.9 &               0.8976   &  3   &               0.4771\\
18     &    6.5 &               0.8129   &  2   &               0.3010\\
20     &    5.44&               0.7356   &  1   &               0.0000\\
\hline
\end{tabular}
\end{center}
\end{table}

\section{GENERAL EQUATION FOR STAR FORMATION}
We suggest, that two fundamental parameters (Elmegreen 1993) which determine star formation may be put in the form
\begin{equation}
R^{n} = \alpha a^{n} + \beta (a f_{P})^{n}
\end{equation}
where $a \equiv 2\sqrt{2} A\sigma_{g}^{2} / Q_{A} \sigma_{c}$ and 
$f_{P} \equiv (P/G \sigma_{c}^{2})^{1/2}$, 
dissipation and $P$ is the pressure. The dimensionless pressure $f_{P}$ 
(defined originally by Elmegreen 1993) is the
square root of the ratio of cloud collision rate to the gravitational 
instability rate and so is a measure of the relative importance of cloud
collisions.  In our discussions presented here, we make use of some interesting results of 
Elmegreen (1993). Both, $Q_{A}$ and $f_{P}$ 
now determine star formation rate. When both are large (i.e., $Q_{A} > 1$ 
and $f_{P}\gg 1$), either thermal instability (macroscopic) triggers star formation 
or cooling which is very effective reduces $Q_{A}$ till gravitational instabilities 
switch over. When both are small ($Q_{A}<1$ and $f_{P}\ll 1$), gravitational 
instabilities form clouds quickly but star formation is hampered due to lack of 
energy dissipation. However, when $Q_{A}\geq 1$ and $f_{P}\ll 1$, star formation proceeds 
via random cloud collisions triggered by thermal instability and rate $R$ is determined 
by second term in equation (15). This is believed to occur 
at galactic radii $r < 4$ kpc and $r > 8$ kpc where one observes $Q\geq 1$ (for an 
observed Q-distribution in the Galaxy, see e.g. WS). When $Q_{A}\leq 1$ 
and $f_{P}\gg 1$, gravitational instability is primarily responsible for both cloud 
and star formation at all radii. In this case, the first term in equation (15) 
determines $R$. It is found that at all radii star formation is governed by relative 
competence of either of these terms. It also becomes evident that virtually 
$Q_{A}< 1$ (or $Q < 1$) is not an absolute criterion for star formation, instead it 
proceeds continuously until requisite ingredients are fuelled in and physical 
conditions are met. In fact, one observes significant star formation even 
when $Q_{A}>1$ in the Galaxy. Thus, the process of star formation can be visualized 
through the equation (15). It may be remarked that our model does not take 
into account the galactic bulge component (Oort 1977) which might contribute 
to galactic gas dynamics in the inner region below 0.1 kpc.

We write star formation rate as
\begin{equation}
R^{n} = \left(\frac{1}{1-\delta}\right)^{n} \frac{d \sigma_{g}^{n}}{dt^{n}},
\end{equation}
where $\delta$ is the fraction of mass returned to interstellar medium from the 
stellar content. From equations (15) and (16) we get
\begin{equation}
\frac{\alpha(2 \sqrt{2} A)^{n} \sigma_{g}^{2n}}{Q_{A}^{n}\sigma_{c}^{n}} + \frac{\beta(2\sqrt{2}A)^{n}\sigma_{g}^{2n}}{Q_{A}^{n}\sigma_{c}^{n}}f^{n}_{P} =
\left(\frac{1}{1-\delta}\right)^{n}\frac{d\sigma_{g}^{n}}{dt^{n}}.
\end{equation}
Assume that parameters $f^{n}_{P}$ and $\sigma_{c}^{n}$ are independent of time. Let us write 
Equation (17) in the form 
\begin{eqnarray}
\left[\alpha(1-\delta)^{n}(2\sqrt{2}A)^{n}+\beta(1-\delta)^{n}(2\sqrt{2}A)^{n} f^{n}_{P}\right]dt^{n} 
\nonumber
 \\ = \frac{(1-f_{c}^{2})^{n/2}}{f_{c}^{2n}}df_{c}^{n}.
\end{eqnarray}
Integrate equation (18) to obtain
\begin{equation}
\frac{t^{n}}{\tau_{\alpha}^{n}} + \frac{t^{n}}{\tau_{\beta}^{n}}f^{n}_{P} = -\frac{(1-f_{c}^{2})^{n/2}}{f_{c}^{n}} -\sin^{-1}(f_{c}^{n}) + constant,
\end{equation}
where we have put
\begin{eqnarray}
\tau_{\alpha}^{-n} &=& \alpha (1-\delta)^{n} (2\sqrt2 A)^{n} \nonumber \\  
\tau_{\beta}^{-n}  &=& \beta (1-\delta)^{n} (2\sqrt2 A)^{n}.     
\end{eqnarray}
If we express $f_{g} = \sigma_{g}/\sigma_{i}$, $f_{ci} = \sigma_{i}/\sigma_{c}$, 
where the subscript i denotes initial values of quantities, we can write equation (19) as
\begin{equation}
\frac{t^{n}}{\tau^{n}_{\alpha}} + \frac{t^n}{\tau_{\beta}^{n}} f^{n}_{P} = - \frac{(1-f_{g}^{2} f_{ci}^{2})^{n/2}}{f_{g}^{n}f_{ci}^{n}} - \sin^{-1} (f_{g}^{n}f_{ci}^{n}) + constant.
\end{equation}
At $t = 0$, $\sigma_{g} = \sigma_{i}$ ($f_{g} = 1$), one obtains the value of 
constant in equation (21) as
\begin{equation}
constant = \frac{(1-f_{ci}^{2})^{n/2}}{f_{ci}^{n}} + \sin^{-1}(f_{ci}^{n}).
\end{equation}
Now, equation (21) becomes 
\begin{eqnarray}
(1+f^{n}_{P})\frac{t^{n}}{\tau^{n}} &=& -\frac{(1-f_{g}^{2}f_{ci}^{2})^{n/2}}{f_{g}^{n}f_{ci}^{n}} \nonumber \\
&+& \frac{(1-f_{ci}^{2})^{n/2}}{f_{ci}^{n}} - \sin^{-1}(f_{g}^{n}f_{ci}^{n}) \nonumber \\
&+& \sin^{-1}(f_{ci}^{n})
\end{eqnarray}
where we have put $\tau_{\alpha}^{n} = \tau_{\beta}^{n} = \tau^{n}$. The contribution of second parameter may be observed on the 
right hand side of equation (23). 
For $f_{P}=0$ and $n=1$, equation (23) reduces (except for a minus sign) to the equation derived by WS (cf. eq.(23) in WS). 
When $f_{g}\ll 1$ near the centre of the disk, we find 
\begin{equation}
\sigma_{g}^{n} \sim \frac{\sigma_{c}^{n}}{1+f^{n}_{P}} \left(\frac{\tau}{t}\right)^{n}.
\end{equation}
Toward the centre, $A$ increases which shows that gas surface density decreases 
($\tau$ varies inversely with $A$). Large values of $f_{P}$ for diffuse clouds again 
guarantees the depletion of gas in the centre. For outer parts of disk, 
$f_{g}\sim 1$, and after expanding various terms in equation (23) and neglecting higher order terms, we get the following 
\begin{equation}
\sigma_{g}^{n} = \sigma_{i}^{n}\left[(1+f^{n}_{P})f_{ci}^{n}\left(\frac{t}{\tau}\right)^{n}+1\right].
\end{equation}
In view of large values of $\tau$ and $f_{ci}\ll 1$, the gas density scales as 
the initial one.

We now proceed to obtain critical column densities based on $\kappa$ and that on new parameter $Q_{A}$. We write
\begin{eqnarray}
\sigma_{cr,\kappa} &=& \frac{\kappa v_{eff}}{\pi G}\nonumber\\
\sigma_{cr,A} &=& \frac{2\sqrt{2} A v_{eff}}{\pi G}\,.
\end{eqnarray} 
Assume a rotation curve of the form $V \propto r^{\mu}$ ($\mu = 0$ for flat
curve). One gets 
\begin{equation}
\frac{\sigma_{cr,A}}{\sigma_{cr,\kappa}}=\frac{1-\mu}{\sqrt{1+\mu}}\,.
\end{equation}
It is found that for $\mu = 0$, both densities agree. But for large $\mu$ (i.e. departures from flatness), 
$\sigma_{cr,A}$ becomes smaller than $\sigma_{cr,\kappa}$. For example, for M 33, $\mu = 0.3$ (Newton 1980) which yields 
\begin{equation}
\sigma_{cr,A} = 0.61 \sigma_{cr,\kappa}\,. 
\end{equation}
Observations of $\sigma_{g}$ (Wilson, Scoville \& Rice 1991) for this galaxy are better explained if one takes $\sigma_{cr,A}$ as threshold density rather
than $\sigma_{cr,\kappa}$ (see also Elmegreen 1993). Thus, $Q_{A}$ emerges as 
a better parameter relating to disk instabilities than Toomre Q-parameter for 
star formation. This is also supported by the ratio of the two threshold densities (see e.g. Table 2).

\begin{table}
\caption{Variation of ratio of threshold densities with index $\mu$}
\begin{center}
\begin{tabular}{cc}
\hline
$\mu$  &  $\sigma_{cr,A}/\sigma_{cr,\kappa}$ \\
\hline                 
               0.005 &               0.99\\
               0.05  &               0.93\\
               0.10  &               0.86\\
               0.15  &               0.80\\
               0.20  &               0.73\\
               0.30  &               0.61\\
               0.40  &               0.51\\
               0.50  &               0.41\\
               0.60  &               0.32\\
               0.70  &               0.23\\
               0.80  &               0.12\\
               0.90  &               0.07\\
\hline
\end{tabular}
\end{center}
\end{table}

Table 2 shows that for highly non-linear region of rotation velocity, threshold density based on $Q_{A}$ is lowered (relatively) favouring 
instability for star formation. On the other hand, threshold density based on Q-parameter is higher (by about one order) in this region. 
This shows that Q-parameter is relatively less efficient to favour star formation. We, therefore, 
conclude that in the non-linear regime of rotation velocity curve, the Q-parameter is approximately $10\%$ 
less effective than $Q_{A}$ parameter for triggering the process.

\begin{table}                             
\caption{Radial variation of $Q_{A}/Q$ for the Galaxy}
\begin{center}
\begin{tabular}{ccccc}
\hline
r &  A  &  --B  & $\kappa$   &  $Q_{A}/Q$ \\
(kpc) & ($\cdot 10^{-16.5} s^{-1}$) & ($\cdot 10^{-16.5} s^{-1}$) & ($\cdot 10^{-16.5} s^{-1}$) &  \\
\hline
1 &       105  &     62   &       203.5  &           1.5\\
2 &       30   &     55   &       136.8  &           0.6\\
3 &       20.9 &     44.5 &       107.9  &           0.5\\
4 &       19.7 &     34.1 &        85.7  &           0.6\\
5 &       19.1 &     26.1 &        68.7  &           0.8\\
6 &       18.2 &     20.1 &        55.5  &           0.9\\
7 &       17.2 &     15.6 &        45.2  &           1.1\\
10&       13.8 &      7.9 &        26.2  &           1.5\\
12&       11.5 &      5.5 &        19.3  &           1.7\\
14&       9.6  &      4.3 &        15.5  &           1.7\\
16&       7.9  &      3.6 &        12.9  &           1.7\\
18&       6.5  &      3.3 &        11.4  &           1.6\\
20&       5.44 &      3.11&        10.3  &           1.5\\
30&       2.91 &      2.40&         7.1  &           1.2\\
50&       1.59 &      1.53&         4.4  &           1.0\\
75&       1.06 &      1.00&          2.9 &           1.0\\
\hline
\end{tabular}
\end{center}
\end{table}

\begin{figure}
\begin{center}
    \leavevmode
    \epsfxsize=8cm
    \epsfysize=8cm
    \epsffile{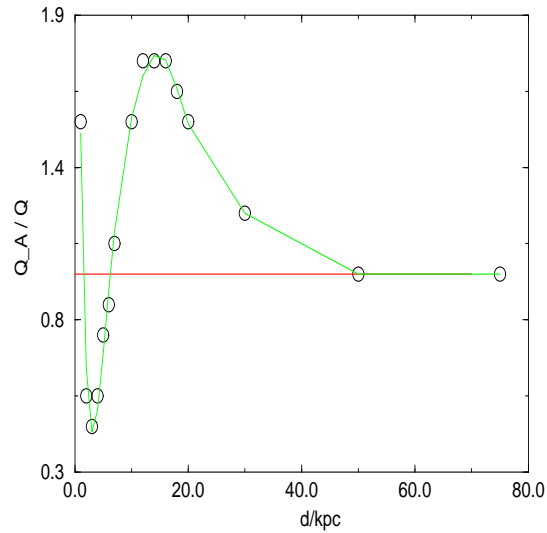}\vspace{0.5cm}
    \caption{Radial variation of $Q_{A}/Q$ for the Galaxy. We have taken
                data from Einasto (1979) and have expressed $Q_{A}$ parameter
                as in Elmegreen (1993).} 
\end{center}
\end{figure}

We have computed the ratio $Q_{A}/Q$ (see e.g. Table 3, data taken from Einasto 1979) at various radial distances of the Galaxy. $B$ denotes the 
second Oort constant. A plot of $Q_{A}/Q$ with radial distances from the centre is shown in Fig. 1. It is found that $Q_{A}$ depicts almost 
similar behaviour as the observed Q-distribution (cf. Fig. 6 in WS) remarkably for the range 1 kpc to 15 kpc and stays 
at $Q_{A}/Q\geq 1$ beyond 30 kpc. Thus, $Q_{A}$ and Q both parameters agree beyond 30 kpc, 
i.e., in the flat rotation curve region. We obtain same result from the data shown in Table 2. 
In fact, for a disk radius below 30 kpc deviations in the two parameters become significant which shows the relative merit of 
$Q_{A}$ parameter over $Q$-parameter to keep track of the physical process like star formation and other nuclear activity as well.

\subsection{COMPARISON OF OBSERVATIONS FOR THE GALAXY}
We assume constant IMF in the solar neighbourhood (Miller \& Scalo 1979; Scalo 1986) 
and take the following input parameters: initial surface density $\sigma_{i,\odot}\simeq \sigma_{g,\odot} + \sigma_{s,\odot} \simeq 50
M_{\odot} pc^{-2}$ (Kuijken \& Gilmore 1989; Bahcall, Flymn \& Gould 1992), $\sigma_{g} \simeq 10 M_{\odot} pc^{-2}$ (McKee 1990), 
$f_{g} \sim 0.2$, $f\sim 0.05$ (Elmegreen 1993), t = age of the Galaxy = 15 Gyr, $\alpha = 0.1$ (Myers et al. 1986), 
Oort shear constant $A = 15$ km s$^{-1}$ kpc$^{-1}$ (Kerr \& Lynden-Bell 1986), $\delta = 0.3$ (Miller \& Scalo 1979; Scalo 1986). 
For $n = 1$, we get timescale of star formation as \mbox{$\tau = 0.38$ Gyr}. 
We calcualte $f_{ci,\odot}$  using equation (23) as $f_{ci,\odot}\sim 0.10$. After substituting these values into equation (11), 
we get star formation rate as $R = 3.8 M_{\odot} pc^{-2} Gyr^{-1}$. This is in agreement with Scalo (1986) who infers 
$R \simeq (1-4) M_{\odot} pc^{-2} Gyr^{-1}$ within an uncertainity factor of about 3. For $n= 2$, we get $\tau = 0.12$  Gyr, $f_{ci,\odot}\sim 0.05$  and 
star formation rate  $R = 6.0 M_{\odot} pc^{-2} Gyr^{-1}$.

We find that our model with $n = 1$ provides star formation rate which is in good agreement with inferred rate in the solar neighbourhood. 
It is to be remarked that our models are sensitive enough to efficiency $\alpha$ introduced in 
equation (1) which however is determined by star formation timescale $\tau$. We note that even if the efficiency drops by $10 \%$ (i.e. when 
the value of $\alpha$ becomes of the order of 0.01) model with $n = 2$ gives same star formation rate as model with $n = 1$ and $\alpha = 0.1$. 
Parametric freedom for $\alpha$ and $f_{P}$ even when $Q_{A} \geq 1$  (i.e. non-gravitational instabilities are dominant) provide a 
general star formation scenario. Our model thus presents a generalisation of WS model with a dependence of star formation 
rate on Oort shear constant $A$. In contrast to WS, we find continuous (in the sense of Q-values) star formation 
rate obeying a similar but however different criterion (i.e. $Q_{A}<1$) of gravitational instability for gaseous disks. In fact, competitive 
nature of two terms in equation (15) helps one to visualize the essence of continuity in the star formation process. 
We discuss the scenario in more details in Sect. 3.3.

\subsection{TIMESCALE OF GAS DEPLETION} 
For a particular $n$, we get from equations (1) and (16) as 
\begin{equation}
\frac{d\sigma_{g}^{n}}{\sigma_{g}^{n}} = -\alpha(1-\delta)^{n} \tau^{-n} dt^{n}\,.
\end{equation}
Integrate equation (29) to obtain 
\begin{equation}
\ln \sigma_{g}^{n}= -\alpha (1-\delta)^{n} \tau^{-n} t^{n} + constant.
\end{equation}
At $t = 0$, $\sigma_{g}^{n}(r,t) = \sigma_{g}^{n}(r,0)$ which yields
\begin{equation}
\sigma_{g}^{n}(r,t) = \sigma_{g}^{n}(r,0) \exp\left[-\alpha (1-\delta)^{n} \tau^{-n} t^{n}\right]
\end{equation}
(see also Lynden-Bell 1975; G\"usten \& Mezger 1983). Now we can write e-folding time as
\begin{equation}
t_{d}^{n} = \frac{1}{\alpha (1-\delta)^{n} \tau^{-n}}\,.
\end{equation}
For our input parameters, the depletion time $t_{d}$ for model $n = 1$ is $t_{d}\simeq 5.4$ Gyr and for model 
$n = 2$, $t_{d} \simeq 0.54$ Gyr. Model with $n = 2$ has $10\%$ depletion time as compared to $n = 1$. For an age of 15 Gyr of the Galaxy model 
with $n = 1$ implies that present gas fraction is $\sim 10 \% $ of its initial value assuming that there is little variation over the last 5 Gyr (Dopita 1985, 1987).

\subsection{THE $f_{P}$-PARAMETER AND STAR FORMATION} 
The $f_{P}$-parameter introduced in equation (15) requires further analysis  as regards the process 
of star formation. It is dimensionless and measures the fraction of diffuse clouds to self-gravitating clouds. 
Low values of $f_{P}$ ($f_{P} \sim 0.01$) means that clouds are dense and self-gravitating. In this case, 
physics of star formation is largely determined by the first term in equation (15). If, however, $f_{P}\sim 100$ as 
for example in the inner Galaxy where pressure becomes high (Elmegreen \& Elmegreen 1987, Polk et al. 1988, see also Vogel, Kulkarni \& 
Scoville 1988 for M 51), as a result diffuse molecular clouds collide and cool leading to large mass 
cloud formation. Nevertheless, this does not mean that such regions result into large star-forming clouds. 
In fact, gravitational instabilities are more efficient (as compared to diffuse cloud collision) to produce 
large mass star-forming clouds. But, then, in this case local energy dissipation occurs through 
diffuse cloud collisions (Elmegreen 1989). A major difficulty for gravitational instability triggered star formation 
appears when $Q_{A}$ and $f_{P}$ both are large. In this case, only thermal instability is responsible for switching 
on star formation process. Murray \& Lin (1989) have stressed the domonating role of thermal instability over gravitational 
instability for a proto-globular cluster where fragmentation (into protostars) is initiated by the former. Low $f_{P}$ values 
may also result when pressure becomes low (i.e. in the outer spiral arms of galaxies where gravity is not significant to 
form large molecular clouds) and star formation proceeds via shear instability. This instability does not depend upon $Q_{A}$. 
Still, $Q_{A}$ has to be relatively small to guarantee unstable radial motion which in turn facilitates dense cloud formation. 
Hence, the new parameter $f_{P}$ appearing in equation (15) also explains star formation scenario plausibly in 
the spiral arms and outer parts of galaxies.

We suggest that two competitive parameters ($Q_{A}$ and $f_{P}$) may be
understood to ultimately decide star formation (in view of observed continuity of the process from 
centre to arm for the Galaxy) as follows. We propose a relation 
\begin{eqnarray}
f_{P} &= 10^{-1-Q_{A}} \hspace{0.5cm}\rm{for}\hspace{0.7cm}Q_{A}<1 \\
f_{P} &= 10^{1+Q_{A}} \hspace{0.7cm}\rm{for}\hspace{0.7cm}Q_{A}\geq 1 \nonumber\,.
\end{eqnarray}
From equations (33), we find that $Q_{A}$ changes continuously from its value
$-(1+\log f_{P})$ to $(-1+\log f_{P})$ respectively from gravitational instability region ($f_{P}\sim 0.01$) to non-gravitational instability region 
($f_{P}\sim 100$) and star formation is supposed to occur from near the centre to outer spiral arms in contrast to earlier suggestion 
(Quirk 1972) who favours (necessarily) gravitational instability to trigger the process.

\section{VARIATION IN THE STAR FORMATION RATE}
We assume cloud mass density in the solar neighbourhood, $\sigma_{c}\sim 100\, M_{\odot}\, pc^{-2}$, 
as constant (Larson 1981). We further assume $f_{c}\sim 0.01$ at $d = 1\, kpc$ (since $\sigma_{g}$ at 1 kpc is $ \sim 100\, M_{\odot}\, pc^{-2}$ 
which makes $f_{c} = 1$ yielding an infinite $R$) to keep star formation reasonably large in the model. Components of surface densities (data taken 
from Einasto 1979; Lacey \& Fall 1985 and WS) are given in Table 4 and plotted as Figure 2 at various distances from the Galactic centre.

\begin{table}
\caption{Radial variation of surface density for the Galaxy}
\begin{center}
\begin{tabular}{ccc}
\hline
d  &log $\sigma_{i}(M_{\odot} pc^{-2})$  &log $\sigma_{g}(M_{\odot} pc^{-2})$\\
(kpc)&     &  \\
\hline
0.001   &           5.5         &                                \\            
0.01    &                       &                                \\         
0.1     &           4.4         &                                \\
1.0     &           3.0         &                          2.0   \\
2.0     &           3.0         &                          0.5   \\
3       &           2.8         &                          0.7   \\
4       &           2.7         &                          1.0   \\
5       &           2.5         &                          1.0   \\
6       &           2.4         &                          1.0   \\     
7       &           2.2         &                          1.0   \\
10      &           1.7         &                          0.9   \\
12      &           1.4         &                          0.7   \\
14      &           1.0         &                          0.6   \\
16      &           0.7         &                          0.5   \\
18      &           0.4         &                          0.3   \\
20      &           0.1         &                          0.0   \\
\hline
\end{tabular}
\end{center}
\end{table}

\begin{figure}
\begin{center}
    \leavevmode
    \epsfxsize=8cm
    \epsfysize=8cm
    \epsffile{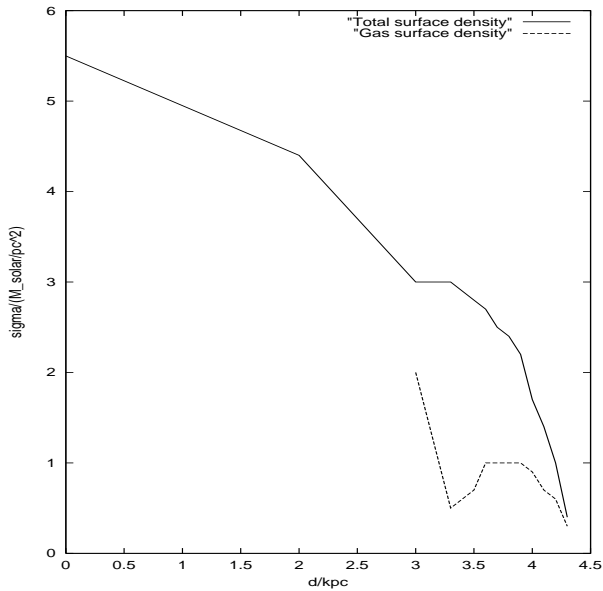}\vspace{0.5cm}
\caption{Radial variation of total surface density $\sigma_{i} 
                (M_{\odot} pc^{-2})$ and gas surface density $\sigma_{g} 
                (M_{\odot} pc^{-2})$ for the Galaxy. The solid line represents  
                the data taken from Lacey \& Fall (1985) and WS; the dotted line 
                represents the data from Einasto (1979).}
\end{center}
\end{figure}

We infer from Figure 2 that the Einasto model shows $\sigma_{i}\sim r^{-0.8}$ dependence for $r\leq 6$ kpc and 
deviates for $r>6\, kpc$ (see also e.g.\ Kundt 1990 for a variant in the Galactic mass distribution). Now, we aim to 
discuss the variation of star formation rate normalised to that in the solar neighbourhood and therefore 
calculate $R/R_{\odot}$ (see e.g.Table 5) using data from Einasto (1979) and plot as 
Figure 3 at various Galactocentric distances. From Figures 2 and 3 we infer that star formation rate 
varies like gas component of surface density.
\begin{table}
\caption{Radial variation of star formation rate for the Galaxy}
\begin{center}
\begin{tabular}{ccccc}
\hline
d &          A  &     $\sigma_{i}$  &      $f_{c}$ &   $R/R_{\odot}$ \\
(kpc)  &   ($\cdot10^{-16.5} s^{-1}$) & ($M_{\odot} pc^{-2 }$) &  & \\ 
\hline
1  &        105    &       1016.3   &            0.01  &      78.6   \\
2  &         30    &        851.1   &            0.03  &      56.5   \\
3  &         20.9  &        633.9   &            0.05  &      48.8   \\
4  &         19.7  &        452.9   &            0.10  &      66.0   \\
5  &         19.1  &        318.4   &            0.11  &      49.5   \\
6  &         18.2  &        222.3   &            0.10  &      29.9   \\
7  &         17.2  &        154.9   &            0.10  &      19.7   \\
10 &         13.8  &         50.9   &            0.07  &       3.6   \\
12 &         11.5  &         23.6   &            0.05  &       1.0   \\
14 &          9.6  &         10.7   &            0.04  &       0.3   \\
16 &          7.9  &          4.9   &            0.03  &       0.1   \\
18 &          6.5  &          2.3   &            0.02  &       0.02  \\
20 &          5.44 &          1.2   &            0.01  &       0.005 \\
\hline
\end{tabular}
\end{center}
\end{table}

\begin{figure}
\begin{center}
    \leavevmode
    \epsfxsize=8cm
    \epsfysize=8cm
    \epsffile{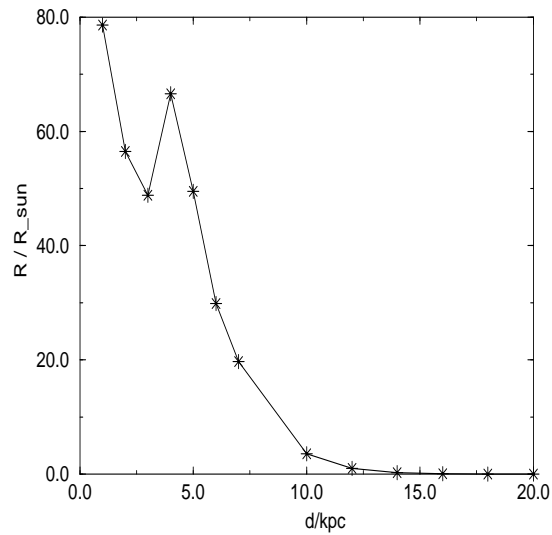}\vspace{0.5cm}
\caption{Star formation rates normalized to its value in the 
         solar neighbourhood. The data are based on Einasto (1979);
         Lacey \& Fall (1985) and  WS.}
\end{center}
\end{figure}

A minimum in $\sigma_{g}$ occurs at $\sim 3$ kpc where we also observe a minimum  in the star formation rate. 
Thereafter, $\sigma_{g}$ increases again and reaches a maximum at $\sim 4$ kpc where we 
observe corresponding increase and maximum in $R/ R_{\odot}$. Our model may be applied to a 
general dependence of $\sigma_{g}$ on the exponent $n$. A given value of n determines the timescale of star formation ($\tau$) 
which however yields the corresponding R. For $n = 1$, our model agrees with WS model but we obtain larger Schmidt exponent (see e.g. equation (12)).

The star formation rates inferred from (i) pulsar data (Lyne, Manchester \& Taylor 1985), (ii) from observations of supernova remnants (Guibert, 
Lequeux \& Viallefond 1978), and (iii) Lyman-continuum photon observations from H II regions (G\"usten \& Mezger 1983) are consistent from our model at all 
radial distances. For example, higher rate of star formation traced by Lyman-continuum near 4 kpc agrees with our model calculations. 
This is evident by the maximum in Figure 3 at 4 kpc from the Galactic centre. In view of comments (Wyse \& Silk 1989) regarding higher star 
formation rates of G\"usten \& Mezger than those given by Scalo (1988) (viz. these estimates may be higher by an order of magnitude) and 
also the fact that it does not match with star formation profile obtained by other techniques (Rana \& Wilkinson 1986), our values are 
apparently better tuned.

We assume $\tau = 0.45$ Gyr. Using parameters as decribed in Section 3.1, we calculate efficiency $\alpha$ of star formation as a function of 
distance from the galactic centre. It is interesting to observe that $\alpha$ changes in the solar neighbourhood. Small values of $\alpha$ at 1 kpc may be 
understood to arise because of shear instability which removes growth of perturbations. Star formation can proceed if the self-gravitational collapse time 
becomes shorter than the shear time ($\sim 0.01$ Gyr). However, relatively large value of $\alpha$ out to 10 kpc does not 
lead to large star formation rate $R/ R_{\odot}$ 
(see Figure 3) due to paucity of gas. In fact, density $\sigma_{i}$ drops below the observed value (Wilson, Scoville \& Rice 1991) of critical 
density at 14 kpc where we expect turn-off of star formation due to gravitational instability. It is also supported by significant depletion of 
gas at this distance (see e.g. Figure 2). 
The striking feature of our result is that $\alpha$ changes in the solar neighbourhood (an efficiency gradient $\sim 0.0057 kpc^{-1}$) by an 
amount $\sim 0.02$. It seems thus natural to think that the efficiency gradient is responsible for radial abundance gradients which are reported in many disk 
galaxies (Edmunds \& Pagel 1984; Diaz \& Tosi 1984; Tosi \& Diaz 1985). The fact that metallicity gradients may be due to changes in efficiency of star 
formation was suggested previously by Lacey \& Fall (1985). We aim to confirm this suggestion from our calculations too. There is hardly a need to invoke radial 
flows (see discussion in Scalo 1988) in this scenario. 

\subsection{METALLICITY GRADIENT VS EFFICIENCY GRADIENT}
Following Pagel \& Patchett (1975) (see also e.g. Pagel \& Edmunds 1981) model of 
chemical evolution of the Galaxy in the solar neighbourhood, we define $\xi$, 
a mass ratio in the form of long-lived stars, and $p$ as the yield of heavy elements which represents mass ejected per unit mass of long-lived stars 
(cf. Searle \& Sargent 1972; Talbot \& Arnett 1973a). For our model $\sigma_{s} = \xi \sigma_{i}$, $\sigma_{g} = (1-\xi)\sigma_{i}$, $\sigma_{s} = {\xi/(1-\xi)} \sigma_{g}$  
and 
\begin{equation}
\frac{d\sigma_{s}}{dt} = -\alpha \left( \frac{\sigma_g}{\tau} \right) = \left( \frac{\xi}{1 - \xi} \frac{d\sigma_g}{dt}\right).
\end{equation}
Equation (34) gives 
\begin{equation}
\frac{d}{dt}\left(\ln \sigma_{g} \right) = -\frac{\alpha(1-\xi)}{\tau \xi}.
\end{equation}
Integration of equation (35) yields
\begin{equation}
\ln \sigma_{g} = -\frac{\alpha(1-\xi)}{\xi}\frac{t}{\tau} + constant.                               
\end{equation}
At $t = 0$, $\sigma_{g} = \sigma_{i}$ hence $constant = \ln \sigma_{i}$.
Thus, equation (36) takes the form
\begin{equation}
\frac{\sigma_{g}}{\sigma_{i}} = (1-\xi) = \exp \left[-\frac{\alpha(1-\xi)}{\xi}\frac{t}{\tau}\right].                
\end{equation}
Therefore,
\begin{equation}
\tau(t) = -\frac{\alpha(1-\xi)}{\xi}\left[\ln (1-\xi)\right]^{-1} t \,.
\end{equation}    

Metallicity $Z$ is expressed as (Pagel \& Patchett 1975)
\begin{equation}
Z=p\ln \left[\frac{1}{1-\xi}\right] = \frac{p\alpha(1-\xi)}{\xi}\frac{t}{\tau}\,.              
\end{equation}

We see that $\tau$ is now a function of time and is given by equation (38). Time evolution of $Z$ may be written as
\begin{equation}   
\frac{dZ}{dt} = \frac{p}{\tau}=\frac{p\xi}{\alpha (1-\xi) [\ln(1-\xi)]^{-1}t}\,.
\end{equation}
From equation (39) we infer that $Z$ varies linearly both with time and efficiency  of star formation. We assume that 
$\xi \simeq 0.8$ (Talbot \& Arnett 1973a) and $p \simeq 0.7 Z_{\odot}$ (Wang \& Silk 1993) to calculate $Z$ in the solar 
neighbourhood. For an efficiency of $\alpha \sim 0.07$ we find $Z\simeq 1.13 Z_{\odot}$ for solar age. This is in agreement with the plot of 
metallicity in the solar neighbourhood by Wyse \& Silk (1989, cf. Figure 2b). The present model thus provides 
time evolution of metallicity which however depends upon efficiency of star formation.

For an efficiency run of $\alpha\sim0.07, 0.08, 0.09, 0.10,$ we find  $Z/Z_{\odot} \simeq 1.23, 1.17, 1.13, 1.09$ respectively which is 
indepedent of Galactic age provided of course the parameters $p$ and $\xi$ do not change with time. In other words, disk aging alters 
$\tau(t)$ such that $t/\tau(t)$ remains constant (for a fixed $\alpha$) and hence there occurs no change in $Z/Z_{\odot}$. 
Our calculations show that metallicity decreases with increase of $\alpha$ at a given age. In the solar neighbourhood, this may be understood
due to paucity of gas favouring relatively low star formation at large distances (see Figure 3) 
and therefore low metal production (see also Friel \& Janes1993). We note that the observed run of metallicity of G-K dwarfs in our Galaxy
is very sensitive to chemical composition of stars of same age (Tinsley 1975).
Janes \& McClure (1972) have suggested enhancement in the dispersion due to chemical 
inhomogeneities in the Galaxy (Talbot \& Arnett 1973b). The structure of Galactic disk and the 
presence of population gradients are given in Ferrini et al. (1994). For a radial 
distribution of abundances in galaxies one may refer to Moll\'a et al. (1996). They 
have also discussed the chemical evolution of solar neighbourhood, see e.g. 
Pardi et al. (1995).

However, as remarked by 
Tinsley (1975) that the observed dispersion (see also Hearnshaw 1972 for
dispersion) in metallicity for stars of same age may result either partly due 
to chemical inhomogeneities (of interstellar medium) or due to causes altogether 
different, essentially favours this analysis. We find  that metallicity dispersion 
for stars of same age may be due to variation of efficiency $\alpha$
through which different sample stars were processed. This confirms the assumption 
of Rana \& Wilkinson (1986) that metallicity dispersion is due to stellar
processing only. It is found that $\alpha$ depends upon star formation rate and
also the gas component of surface density $\sigma_{g}$. We conclude that 
$\alpha$ predominantly determines the observed dispersion and plays a key role
towards metal enrichment or otherwise of the interstallar medium.

At various disk ages (at a given radial distance), there occurs change in $\alpha$ which causes metallicity dispersion. We notice that $\alpha$ 
also changes at various distances from the Galactic centre which results in spatial metallicity gradients. One immediately finds that apparent metallicity 
dispersions with either age or distance depend upon $\alpha$. The $[O/H]$ vs age plot (Wyse \& Silk 19879, cf. Figure 2d; see also Carlberg et al. 1985) 
shows hardly significant metallicity gradient at all disk ages (cf. Friel \& Janes 1993). We suggest that all sample stars might have 
followed evolution with almost similar efficiency. Thus, the important result of this analysis is the confirmation of the suggestion by Lacey \& Fall (1985) 
and Richtler (1995) regarding metallicity gradients.  For a comprehensive treatment of radial abundance gradients in spiral disks and age-metallicity relation in 
different stellar populations one may refer to Edvardsson et al. (1993); Pagel (1994). An interesting modern analysis of kinematics and abundance distribution 
for our Galaxy has been given by Gilmore, Wyse \& Kuijken (1989). Matteucci (1996) has 
reviewed exhaustively the evolution of the abundances of heavy elements in gas and stars 
(indicating observational and theoretical constraints) in galaxies of different morphological 
types. After a similar work by Tinsley (1980), the article provide a good document of the 
progress in the understanding of the physical processes regulating the chemical evolution 
of galaxies. Formation and evolution of our Galaxy is also discussed. For a review on 
abundance ratios and Galactic chemical evolution see McWilliam (1997). Chemical 
evolution of solar neighbourhood according to the standard infall model using SN II data  
are summarized in Thomas et al. (1998).

\section{DISCUSSIONS} 
We have studied that the suggestion by Elmegreen (1993) regarding star formation 
appears more robust than $Q$-criterion. 
It is because unless $Q\leq 1$, gravitational instability does not permit star 
formation. However, when $Q>1$, the system becomes 
gravitationally stable and consequently star formation via large cloud formation 
is not feasible. A natural question which one 
would ask is: how does star formation proceed when $Q$ enters the stable regime? 
This infact led to an alternative criterion for cloud 
formation (discussed in the text) leading to star formation as originally suggested 
by Elmegreen (1993). Accordingly, when magnetic 
field is taken into consideration, velocity dispersion changes and thus $Q$ is pushed to the stable regime. At this stage, 
non-gravitational instabilities (e.g. thermal instability, shear instability) dominate over gravitational instability. We infer from Figure 1 
that dependence of $Q_{A}/Q$ on distance from the Galactic centre describes the relative merit of $Q_{A}$-parameter over $Q$-parameter beyond 6 kpc. 
Observations of $\sigma_{g}$ for M 33 are in better agreement with theory when one regards $Q_{A}$ as the 
gravitational instability parameter (see e.g. Wilson, Scoville \& Rice 1991). 
It is found that both $Q_{A}$ and $Q$ parameters agree beyond 30 kpc.

We have obtained generalised version of WS equation (see e.g. Sect. 2 equation (12) 
and Sect. 3 equation (23)) in the sense that (i) there is an 
additional non-zero term in equation (12) and (ii) in view of equation (15), one arrives at a natural rescue from cut-off criterion for star formation. 
We also show that the nominal Schmidt exponent $n_{s}$ is given by $2<n_{s}<3$  in our model. We suggest a general equation (e.g. equation (15)) for the 
star formation rate consisting of two terms: the first term dominates when $Q_{A}<1$ and $f_{P}\ll 1$; the second term dominates when $Q_{A}\geq 1$ and $f_{P}\ll 1$. 
Apparently, the relative competence of either of these terms determines star formation scenario as discussed in Sect. 3 at all radial distances. 
Virtually, $Q<1$ (or $Q_{A}<1$) is not an absolute criterion for star formation. For our model with $n = 1$, we get star formation rates which are in good agreement with
values inferred by Scalo (1986). We find that our models are sensitive enough to efficiency $\alpha$ and timescale $\tau$ of star formation. 
A given exponent $n$ determines $\tau$ which however yields the corresponding star formation rate $R$.

We suggest that essentially the efficiency gradient is the cause for radial abundance gradients which are reported in many disk galaxies. 
Under the approximation of closed box model, we have derived time evolution of $\tau(t)$ and also the metallicity Z(t). Both $\tau(t)$ and Z(t) 
are functions of $\alpha$, $p$ (the yield of heavy element) and mass ratio $\xi$. We notice hardly any metallicity change as disk ages which however 
reflects that stellar processing occurs at a fixed $\alpha$. Metallicity dispersion for stars of same age may be caused due to variation in $\alpha$. 
We conclude that $\alpha$ is predominantly responsible for metallicity dispersion and also for metal enrichment of interstellar medium. A simple model 
as discussed above provides some important characteristics of our Galactic disk. Although, as suggested by Tinsley (1980), the star formation is a 
complicated function of several physical parameters, e.g. gas density, gas sound speed, shock frequency, shock strength, gas rotation, shear constant $A$, 
magnetic field, gas metal abundance, and background star density. It is however difficult to predict the actual dependence of $R$ on these parameters. 
One therefore studies some form of R and its consequent effect on chemical and photometric evolution. Finally, the model predictions are compared with observations.

We note that star formation rate was probably higher in the central part of the disk of our Galaxy at an early epoch of evolution. It is to be remarked that 
hydrodynamical simulations of the formation and evolution of a galaxy may be performed incorporating our model formulation of star formation rate and metallicity. 
Model predictions when compared with observations of other galaxies would speak of its robustness and proof.

\vspace{1cm}
\noindent{\bf ACKNOWLEDGEMENTS:} One of us (USP) is thankful to Indo-German Bilateral 
Cooperation Office, Forschungszentrum, J\"ulich for financial support to visit Bonn 
and collaborate with Prof. Dr. Wolfgang Kundt. USP is indebted to Prof. Dr. Kundt for 
exposing to current problems. Thanks are due to Prof. Dr. Joseph Silk for kindly sending 
some of his recent reprints and especially the referee, Prof. Bruce Elmegreen, for helpful 
comments. USP is also thankful to IAU Comm. 38 for travel support from 
the Exchange of Astronomers Programme.\\

\end{document}